


\message
{JNL.TEX version 0.92 as of 6/9/87.  Report bugs and problems to Doug Eardley.}

\catcode`@=11
\expandafter\ifx\csname inp@t\endcsname\relax\let\inp@t=\input
\def\input#1 {\expandafter\ifx\csname #1IsLoaded\endcsname\relax
\inp@t#1%
\expandafter\def\csname #1IsLoaded\endcsname{(#1 was previously loaded)}
\else\message{\csname #1IsLoaded\endcsname}\fi}\fi
\catcode`@=12



\font\twelverm=cmr10 scaled 1200    \font\twelvei=cmmi10 scaled 1200
\font\twelvesy=cmsy10 scaled 1200   \font\twelveex=cmex10 scaled 1200
\font\twelvebf=cmbx10 scaled 1200   \font\twelvesl=cmsl10 scaled 1200
\font\twelvett=cmtt10 scaled 1200   \font\twelveit=cmti10 scaled 1200
\font\twelvesc=cmcsc10 scaled 1200  \font\twelvesf=amssmc10 scaled 1200
\skewchar\twelvei='177   \skewchar\twelvesy='60


\def\twelvepoint{\normalbaselineskip=12.4pt plus 0.1pt minus 0.1pt
  \abovedisplayskip 12.4pt plus 3pt minus 9pt
  \belowdisplayskip 12.4pt plus 3pt minus 9pt
  \abovedisplayshortskip 0pt plus 3pt
  \belowdisplayshortskip 7.2pt plus 3pt minus 4pt
  \smallskipamount=3.6pt plus1.2pt minus1.2pt
  \medskipamount=7.2pt plus2.4pt minus2.4pt
  \bigskipamount=14.4pt plus4.8pt minus4.8pt
  \def\rm{\fam0\twelverm}          \def\it{\fam\itfam\twelveit}%
  \def\sl{\fam\slfam\twelvesl}     \def\bf{\fam\bffam\twelvebf}%
  \def\mit{\fam 1}                 \def\cal{\fam 2}%
  \def\sc{\twelvesc}		   \def\tt{\twelvett}
  \def\sf{\twelvesf}
  \textfont0=\twelverm   \scriptfont0=\tenrm   \scriptscriptfont0=\sevenrm
  \textfont1=\twelvei    \scriptfont1=\teni    \scriptscriptfont1=\seveni
  \textfont2=\twelvesy   \scriptfont2=\tensy   \scriptscriptfont2=\sevensy
  \textfont3=\twelveex   \scriptfont3=\twelveex  \scriptscriptfont3=\twelveex
  \textfont\itfam=\twelveit
  \textfont\slfam=\twelvesl
  \textfont\bffam=\twelvebf \scriptfont\bffam=\tenbf
  \scriptscriptfont\bffam=\sevenbf
  \normalbaselines\rm}



\def\beginlinemode{\endmode
  \begingroup\parskip=0pt \obeylines\def\\{\par}\def\endmode{\par\endgroup}}
\def\beginparmode{\endmode
  \begingroup \def\endmode{\par\endgroup}}
\let\endmode=\par
{\obeylines\gdef\
{}}
\def\singlespace{\baselineskip=\normalbaselineskip}

\def\oneandahalfspace{\baselineskip=\normalbaselineskip
  \multiply\baselineskip by 3 \divide\baselineskip by 2}
\def\doublespace{\baselineskip=\normalbaselineskip \multiply\baselineskip by 2}

\newcount\firstpageno
\firstpageno=2
\footline={\ifnum\pageno<\firstpageno{\hfil}\else{\hfil\twelverm\folio\hfil}\fi}
\def\toppageno{\global\footline={\hfil}\global\headline
  ={\ifnum\pageno<\firstpageno{\hfil}\else{\hfil\twelverm\folio\hfil}\fi}}
\let\rawfootnote=\footnote		
\def\footnote#1#2{{\rm\singlespace\parindent=0pt\parskip=0pt
  \rawfootnote{#1}{#2\hfill\vrule height 0pt depth 6pt width 0pt}}}
\def\raggedcenter{\leftskip=4em plus 12em \rightskip=\leftskip
  \parindent=0pt \parfillskip=0pt \spaceskip=.3333em \xspaceskip=.5em
  \pretolerance=9999 \tolerance=9999
  \hyphenpenalty=9999 \exhyphenpenalty=9999 }
\def\dateline{\rightline{\ifcase\month\or
  January\or February\or March\or April\or May\or June\or
  July\or August\or September\or October\or November\or December\fi
  \space\number\year}}
\def\received{\vskip 3pt plus 0.2fill
 \centerline{\sl (Received\space\ifcase\month\or
  January\or February\or March\or April\or May\or June\or
  July\or August\or September\or October\or November\or December\fi
  \qquad, \number\year)}}


\hsize=6.5truein
\hoffset=0truein
\vsize=8.9truein
\voffset=0truein
\parskip=\medskipamount
\def\\{\cr}
\twelvepoint		
\doublespace		
\overfullrule=0pt	




\def\title			
  {\null\vskip 3pt plus 0.2fill
   \beginlinemode \doublespace \raggedcenter \bf}

\def\author			
  {\vskip 3pt plus 0.2fill \beginlinemode
   \singlespace \raggedcenter\sc}

\def\affil			
  {\vskip 3pt plus 0.1fill \beginlinemode
   \oneandahalfspace \raggedcenter \sl}

\def\abstract			
  {\vskip 3pt plus 0.3fill \beginparmode
   \oneandahalfspace ABSTRACT: }

\def\endtitlepage		
  {\endpage			
   \body}
\let\endtopmatter=\endtitlepage

\def\body			
  {\beginparmode}		

\def\head#1{			
  \goodbreak\vskip 0.5truein	
  {\immediate\write16{#1}
   \raggedcenter \uppercase{#1}\par}
   \nobreak\vskip 0.25truein\nobreak}

\def\beginitems{
\par\medskip\bgroup\def\i##1 {\item{##1}}\def\ii##1 {\itemitem{##1}}
\leftskip=36pt\parskip=0pt}
\def\enditems{\par\egroup}

\def\beneathrel#1\under#2{\mathrel{\mathop{#2}\limits_{#1}}}

\def\refto#1{$^{#1}$}		

\def\references			
  {\head{References}		
   \beginparmode
   \frenchspacing \parindent=0pt \leftskip=1truecm
   \parskip=8pt plus 3pt \everypar{\hangindent=\parindent}}

\def\referencesnohead   	
  {                     	
   \beginparmode
   \frenchspacing \parindent=0pt \leftskip=1truecm
   \parskip=8pt plus 3pt \everypar{\hangindent=\parindent}}

\gdef\refis#1{\item{#1.\ }}			

\gdef\journal#1, #2, #3, 1#4#5#6{		
    {\sl #1~}{\bf #2}, #3 (1#4#5#6)}		

\def\pr{\journal Phys. Rev., }

\def\prb{\journal Phys. Rev. B, }

\def\prl{\journal Phys. Rev. Lett., }

\def\np{\journal Nucl. Phys., }

\def\endreferences{\body}

\def\figurecaptions		
  {\endpage
   \beginparmode
   \head{Figure Captions}
}

\def\endpage			
  {\vfill\eject}

\def\endpaper			
  {\endmode\vfill\supereject}


\def\heading				
  {\vskip 0.5truein plus 0.1truein	
   \beginparmode \def\\{\par} \parskip=0pt \singlespace \raggedcenter}

\def\subheading				
  {\vskip 0.25truein plus 0.1truein	
   \beginlinemode \singlespace \parskip=0pt \def\\{\par}\raggedcenter}

\def\tag#1$${\eqno(#1)$$}

\def\align#1$${\eqalign{#1}$$}

\def\aligntag#1$${\gdef\tag##1\\{&(##1)\cr}\eqalignno{#1\\}$$
  \gdef\tag##1$${\eqno(##1)$$}}

\def\endaligntag{}

\def\overset #1\to#2{{\mathop{#2}\limits^{#1}}}
\def\underset#1\to#2{{\let\next=#1\mathpalette\undersetpalette#2}}
\def\undersetpalette#1#2{\vtop{\baselineskip0pt
\ialign{$\mathsurround=0pt #1\hfil##\hfil$\crcr#2\crcr\next\crcr}}}


\def\ref#1{Ref.~#1}			
\def\Ref#1{Ref.~#1}			
\def\[#1]{[\cite{#1}]}
\def\cite#1{{#1}}
\def\(#1){(\call{#1})}
\def\call#1{{#1}}
\def\taghead#1{}
\def\frac#1#2{{#1 \over #2}}

\def\12{{1\over2}}

\def\ie{{\it i.e.,\ }}

\def\etal{{\it et al.\ }}

\def\sla{\raise.15ex\hbox{$/$}\kern-.57em}
\def\leaderfill{\leaders\hbox to 1em{\hss.\hss}\hfill}
\def\twiddle{\lower.9ex\rlap{$\kern-.1em\scriptstyle\sim$}}
\def\bigtwiddle{\lower1.ex\rlap{$\sim$}}
\def\gtwid{\mathrel{\raise.3ex\hbox{$>$\kern-.75em\lower1ex\hbox{$\sim$}}}}
\def\ltwid{\mathrel{\raise.3ex\hbox{$<$\kern-.75em\lower1ex\hbox{$\sim$}}}}
\def\square{\kern1pt\vbox{\hrule height 1.2pt\hbox{\vrule width 1.2pt\hskip 3pt
   \vbox{\vskip 6pt}\hskip 3pt\vrule width 0.6pt}\hrule height 0.6pt}\kern1pt}
\def\tdot#1{\mathord{\mathop{#1}\limits^{\kern2pt\ldots}}}

\def\pmb#1{\setbox0=\hbox{#1}%
  \kern-.025em\copy0\kern-\wd0
  \kern  .05em\copy0\kern-\wd0
  \kern-.025em\raise.0433em\box0 }

\catcode`@=11
\newcount\r@fcount \r@fcount=0
\newcount\r@fcurr
\immediate\newwrite\reffile
\newif\ifr@ffile\r@ffilefalse
\def\w@rnwrite#1{\ifr@ffile\immediate\write\reffile{#1}\fi\message{#1}}

\def\writer@f#1>>{}
\def\referencefile{
  \r@ffiletrue\immediate\openout\reffile=\jobname.ref%
  \def\writer@f##1>>{\ifr@ffile\immediate\write\reffile%
    {\noexpand\refis{##1} = \csname r@fnum##1\endcsname = %
     \expandafter\expandafter\expandafter\strip@t\expandafter%
     \meaning\csname r@ftext\csname r@fnum##1\endcsname\endcsname}\fi}%
  \def\strip@t##1>>{}}

\def\citeall#1{\xdef#1##1{#1{\noexpand\cite{##1}}}}
\def\cite#1{\each@rg\citer@nge{#1}}	

\def\each@rg#1#2{{\let\thecsname=#1\expandafter\first@rg#2,\end,}}
\def\first@rg#1,{\thecsname{#1}\apply@rg}	
\def\apply@rg#1,{\ifx\end#1\let\next=\relax
\else,\thecsname{#1}\let\next=\apply@rg\fi\next}

\def\citer@nge#1{\citedor@nge#1-\end-}	
\def\citer@ngeat#1\end-{#1}
\def\citedor@nge#1-#2-{\ifx\end#2\r@featspace#1 
  \else\citel@@p{#1}{#2}\citer@ngeat\fi}	
\def\citel@@p#1#2{\ifnum#1>#2{\errmessage{Reference range #1-#2\space is bad.}%
    \errhelp{If you cite a series of references by the notation M-N, then M and
    N must be integers, and N must be greater than or equal to M.}}\else%
 {\count0=#1\count1=#2\advance\count1
by1\relax\expandafter\r@fcite\the\count0,%
  \loop\advance\count0 by1\relax
    \ifnum\count0<\count1,\expandafter\r@fcite\the\count0,%
  \repeat}\fi}

\def\r@featspace#1#2 {\r@fcite#1#2,}	
\def\r@fcite#1,{\ifuncit@d{#1}
    \newr@f{#1}%
    \expandafter\gdef\csname r@ftext\number\r@fcount\endcsname%
                     {\message{Reference #1 to be supplied.}%
                      \writer@f#1>>#1 to be supplied.\par}%
 \fi%
 \csname r@fnum#1\endcsname}
\def\ifuncit@d#1{\expandafter\ifx\csname r@fnum#1\endcsname\relax}%
\def\newr@f#1{\global\advance\r@fcount by1%
    \expandafter\xdef\csname r@fnum#1\endcsname{\number\r@fcount}}

\let\r@fis=\refis			
\def\refis#1#2#3\par{\ifuncit@d{#1}
   \newr@f{#1}%
   \w@rnwrite{Reference #1=\number\r@fcount\space is not cited up to now.}\fi%
  \expandafter\gdef\csname r@ftext\csname r@fnum#1\endcsname\endcsname%
  {\writer@f#1>>#2#3\par}}

\def\ignoreuncited{
   \def\refis##1##2##3\par{\ifuncit@d{##1}%
     \else\expandafter\gdef\csname r@ftext\csname
r@fnum##1\endcsname\endcsname%
     {\writer@f##1>>##2##3\par}\fi}}

\def\r@ferr{\endreferences\errmessage{I was expecting to see
\noexpand\endreferences before now;  I have inserted it here.}}
\let\r@ferences=\references
\def\references{\r@ferences\def\endmode{\r@ferr\par\endgroup}}

\let\endr@ferences=\endreferences
\def\endreferences{\r@fcurr=0
  {\loop\ifnum\r@fcurr<\r@fcount
    \advance\r@fcurr by 1\relax\expandafter\r@fis\expandafter{\number\r@fcurr}%
    \csname r@ftext\number\r@fcurr\endcsname%
  \repeat}\gdef\r@ferr{}\endr@ferences}


\let\r@fend=\endpaper\gdef\endpaper{\ifr@ffile
\immediate\write16{Cross References written on []\jobname.REF.}\fi\r@fend}

\catcode`@=12

\citeall\refto		
\citeall\ref		%
\citeall\Ref		%

\catcode`@=11
\newcount\tagnumber\tagnumber=0

\immediate\newwrite\eqnfile
\newif\if@qnfile\@qnfilefalse
\def\write@qn#1{}
\def\writenew@qn#1{}
\def\w@rnwrite#1{\write@qn{#1}\message{#1}}
\def\@rrwrite#1{\write@qn{#1}\errmessage{#1}}

\def\taghead#1{\gdef\t@ghead{#1}\global\tagnumber=0}
\def\t@ghead{}

\expandafter\def\csname @qnnum-3\endcsname
  {{\t@ghead\advance\tagnumber by -3\relax\number\tagnumber}}
\expandafter\def\csname @qnnum-2\endcsname
  {{\t@ghead\advance\tagnumber by -2\relax\number\tagnumber}}
\expandafter\def\csname @qnnum-1\endcsname
  {{\t@ghead\advance\tagnumber by -1\relax\number\tagnumber}}
\expandafter\def\csname @qnnum0\endcsname
  {\t@ghead\number\tagnumber}
\expandafter\def\csname @qnnum+1\endcsname
  {{\t@ghead\advance\tagnumber by 1\relax\number\tagnumber}}
\expandafter\def\csname @qnnum+2\endcsname
  {{\t@ghead\advance\tagnumber by 2\relax\number\tagnumber}}
\expandafter\def\csname @qnnum+3\endcsname
  {{\t@ghead\advance\tagnumber by 3\relax\number\tagnumber}}

\def\equationfile{%
  \@qnfiletrue\immediate\openout\eqnfile=\jobname.eqn%
  \def\write@qn##1{\if@qnfile\immediate\write\eqnfile{##1}\fi}
  \def\writenew@qn##1{\if@qnfile\immediate\write\eqnfile
    {\noexpand\tag{##1} = (\t@ghead\number\tagnumber)}\fi}
}

\def\callall#1{\xdef#1##1{#1{\noexpand\call{##1}}}}
\def\call#1{\each@rg\callr@nge{#1}}

\def\each@rg#1#2{{\let\thecsname=#1\expandafter\first@rg#2,\end,}}
\def\first@rg#1,{\thecsname{#1}\apply@rg}
\def\apply@rg#1,{\ifx\end#1\let\next=\relax%
\else,\thecsname{#1}\let\next=\apply@rg\fi\next}

\def\callr@nge#1{\calldor@nge#1-\end-}
\def\callr@ngeat#1\end-{#1}
\def\calldor@nge#1-#2-{\ifx\end#2\@qneatspace#1 %
  \else\calll@@p{#1}{#2}\callr@ngeat\fi}
\def\calll@@p#1#2{\ifnum#1>#2{\@rrwrite{Equation range #1-#2\space is bad.}
\errhelp{If you call a series of equations by the notation M-N, then M and
N must be integers, and N must be greater than or equal to M.}}\else%
 {\count0=#1\count1=#2\advance\count1
by1\relax\expandafter\@qncall\the\count0,%
  \loop\advance\count0 by1\relax%
    \ifnum\count0<\count1,\expandafter\@qncall\the\count0,%
  \repeat}\fi}

\def\@qneatspace#1#2 {\@qncall#1#2,}
\def\@qncall#1,{\ifunc@lled{#1}{\def\next{#1}\ifx\next\empty\else
  \w@rnwrite{Equation number \noexpand\(>>#1<<) has not been defined yet.}
  >>#1<<\fi}\else\csname @qnnum#1\endcsname\fi}

\let\eqnono=\eqno
\def\eqno(#1){\tag#1}
\def\tag#1$${\eqnono(\displayt@g#1 )$$}

\def\aligntag#1\endaligntag
  $${\gdef\tag##1\\{&(##1 )\cr}\eqalignno{#1\\}$$
  \gdef\tag##1$${\eqnono(\displayt@g##1 )$$}}

\def\eqalignno#1{\displ@y \tabskip\centering
  \halign to\displaywidth{\hfil$\displaystyle{##}$\tabskip\z@skip
    &$\displaystyle{{}##}$\hfil\tabskip\centering
    &\llap{$\displayt@gpar##$}\tabskip\z@skip\crcr
    #1\crcr}}

\def\displayt@gpar(#1){(\displayt@g#1 )}

\def\displayt@g#1 {\rm\ifunc@lled{#1}\global\advance\tagnumber by1
        {\def\next{#1}\ifx\next\empty\else\expandafter
        \xdef\csname @qnnum#1\endcsname{\t@ghead\number\tagnumber}\fi}%
  \writenew@qn{#1}\t@ghead\number\tagnumber\else
        {\edef\next{\t@ghead\number\tagnumber}%
        \expandafter\ifx\csname @qnnum#1\endcsname\next\else
        \w@rnwrite{Equation \noexpand\tag{#1} is a duplicate number.}\fi}%
  \csname @qnnum#1\endcsname\fi}

\def\ifunc@lled#1{\expandafter\ifx\csname @qnnum#1\endcsname\relax}

\let\@qnend=\end\gdef\end{\if@qnfile
\immediate\write16{Equation numbers written on []\jobname.EQN.}\fi\@qnend}

\catcode`@=12


\def\ie{{\it i.e.,\ }}
\def\etal{{\it et al.}}

\def\>{\rangle}
\def\<{\langle}
\def\o{\over}

\def\t{\tilde}

\def\slD{\raise.15ex\hbox{$/$}\kern-.57em\hbox{$D$}}
\def\dsl{\raise.15ex\hbox{$/$}\kern-.57em\hbox{$\Delta$}}
\def\slp{{\raise.15ex\hbox{$/$}\kern-.57em\hbox{$\partial$}}}
\def\nsl{\raise.15ex\hbox{$/$}\kern-.57em\hbox{$\nabla$}}
\def\sla{\raise.15ex\hbox{$/$}\kern-.57em\hbox{$\rightarrow$}}
\def\slla{\raise.15ex\hbox{$/$}\kern-.57em\hbox{$\lambda$}}
\def\slb{\raise.15ex\hbox{$/$}\kern-.57em\hbox{$b$}}
\def\lnp{\raise.15ex\hbox{$/$}\kern-.57em\hbox{$p$}}
\def\lnk{\raise.15ex\hbox{$/$}\kern-.57em\hbox{$k$}}
\def\lnK{\raise.15ex\hbox{$/$}\kern-.57em\hbox{$K$}}
\def\lnq{\raise.15ex\hbox{$/$}\kern-.57em\hbox{$q$}}

\def\a{\alpha}
\def\be{{\beta}}
\def\ga{{\gamma}}
\def\de{{\delta}}
\def\eps{{\epsilon}}

\def\th{{\theta}}

\def\si{{\sigma}}

\def\om{{\omega}}
\def\Ga{{\Gamma}}

\def\De{{\Delta}}
\def\Th{{\Theta}}

\def\Om{{\Omega}}

\def\cT{{\cal T}}

\def\part{\partial}

\def\dag{\dagger}

\def\abs{
         \vskip 3pt plus 0.3fill\beginparmode
         \doublespace ABSTRACT:\ }

\input epsf
\singlespace
\def\r{\rho}
\def\eps{E}

\title Impurity effects on chiral 1D electron systems

\author Xiao-Gang Wen

\affil
Department of Physics, MIT
77 Massachusetts Avenue
Cambridge, MA 02139, U.S.A.

\abs{We studied low energy effects of impurities in several
chiral 1D electron systems that contain only excitations moving in one
direction. We first considered single-impurity scattering between two
branches of 1D chiral Luttinger liquids.
The general form of impurity scattering
matrix was found which, some times, allow the chemical
potentials on the outgoing branches to be higher (or lower) then the maximum
(or minimum) chemical potentials on the incoming branches. We also studied
the effects of many impurities (with finite density) on two branches of 1D
chiral Fermi liquids (that appear on the edge of $\nu=2$ integral
quantum Hall liquid). We
found that the impurities drive the system to a new fix point at low energies
which has an $SU(2)$ symmetry even when the two branches originally
have different velocities. (Note a clean
system is $SU(2)$ symmetric only when the two branches
have the same velocity.) The probability distribution and correlation of
scattering matrices at different energies are described by universal
functions which are calculated exactly for this new fixed point.
}

\endtopmatter

\head{1. Introduction}

It is well known that impurities have drastic effects on transport
properties of 1D electron systems (which contain both right and left
moving excitations).
Electrons will be localized no matter how weak is the impurity potential.
It is also well known that impurities can not localize chiral 1D electron
systems (which contain only excitations that move in one direction)
due to the lack of back scattering, or more generally, due to the gauge
invariance.\refto{Halp,edge,edgere} The gauge invariance insures the electrons
to remain delocalized in presence of impurities even when the back propagating
modes do exist, such as on the edge of the $\nu=2/3$ fraction quantum Hall
liquid.\refto{elec,JM,edgere}
Chiral 1D electron systems appear on edges of quantum Hall (QH) liquids.
Absence of localization in chiral systems has important consequences on edge
transport of QH liquids.\refto{Halp}

Many HQ liquids contain several branches of edge
excitations.\refto{M,BW,FK,edgere} Impurities
near the edges may scatter electrons or quasiparticles from one branch
to another (or equivalently induce tunneling between different branches).
In some cases (depending on the internal structures, \ie the topological
orders,\refto{top} of the the bulk QH liquids and interactions along the edge),
one can show that the impurity scattering is relevant
and becames strong at low energies. Thus one expects impurities have some
drastic effects at low energies. For chiral systems impurities cannot induce
localization. One may wonder what kind of effects that impurities may have
at low energies.

In this paper we study low energy effects of impurities in several
simple chiral electron systems. In sections 2 and 3 we study the effects of a
single-impurity. Here we assume that the edge branches are described by
chiral Luttinger liquids.\refto{edgere}
 A general form of impurity scattering
matrix is obtained
which depends on a single parameter that characterize the strength
of the impurity scattering. We find surprisingly that for strong enough
scattering the chemical
potentials on the outgoing branches can be higher (or lower) then the maximum
(or minimum) chemical potential on the incoming branches. In some special
cases the tunneling between chiral Luttinger liquids can be mapped into a
problem of tunneling between two Fermi liquids. This mapping
 allows us to apply the
results for the tunneling between Fermi liquids to the tunneling between the
Luttinger liquids in these special cases, which include the tunneling
between two branches of the edge states of $\nu=2/5$ fractional QH (FQH)
 liquid. The mapping
allows us to obtain noise spectrum for the tunneling current. From the noise
spectrum one can directly measure the fractional charges that appear in the
FQH liquids.

We also study (in sections 3 and 4)
the effects of many impurities (with finite density) in a system that
contains two branches of 1D
chiral Fermi liquids. (Such a system appears on the edge of $\nu=2$ QH
liquid\refto{Halp}). We
find that the impurities drive the system to a new fix point at low energies
which has an $SU(2)\times SU(2)$ symmetry
even when the two branches originally have different velocities.
(Note the pure system has an $SU(2)$
symmetry only when the two branches
have the same velocity.) At this new fixed point the probability
distribution and correlation of
scattering matrices are described by universal functions which are
calculated exactly at zero and
finite temperatures. We also calculated the exact
impurity averaged
density-density correlation function at finite temperatures.
Our results can be easily generalized
to systems that contain $n$ branches of chiral 1D Fermi liquids
where the infrared fix point will has an
$SU(n)\times SU(n)$ symmetry.

\head{2. Scattering between chiral Luttinger liquids
 through a single impurity --
general considerations}

Consider a boundary between two FQH liquids or between a
FQH liquid and vacuum. Assume that the boundary contains two branches of
edge excitations, and that
the conductances of the
two FQH edge branches are given by $\si_i=\nu_i e^2/h$, $i=1,2$.
The electrons and/or quasiparticles can be scattered by an
impurity from one edge to the other. (see Fig. 1)
(The situation with
$\nu_1=\nu_2=1$
has been discussed in \Ref{FH}.) The scattering configuration in Fig. 1
can be realized in experimental devices in several different ways as
illustrated in Fig. 2. Let $I_a$, $a=1,2$ ($a=3,4$), be the incoming
(outgoing)
currents on the edge 1 and 2. Let $\mu_a$, $a=1,2$, be chemical potentials
on the two incoming branches. We have
$$I_a=\nu_a {e\o h}\mu_a,\ \ \ \ \ \ \ a=1,2
\eqno(0.1)$$
In the linear response regime, the outgoing
currents and the incoming currents are related through
$$\pmatrix{I_3\cr I_4\cr}=\pmatrix{a&b\cr c&d\cr}
\pmatrix{I_1\cr I_2\cr}\equiv B\pmatrix{I_1\cr I_2\cr}
\eqno(0.2)$$
where $B$ will be called the branching matrix.
The current conservation requires the branching matrix to satisfy that
$$ a+c=1,\ \ \ \ \ \ \ \ b+d=1.
\eqno(0.3)$$
Another condition on the branching
matrix $B$ can be obtained by noticing that when
$\mu_1=\mu_2$ there is no net current passing between the two edges and
hence $I_1=I_3$ and $I_2=I_4$. Mathematically this implies that
$\pmatrix{\nu_1\cr \nu_2\cr}$ is an eigenvector of $B$:
$$B\pmatrix{\nu_1\cr \nu_2\cr}=\pmatrix{\nu_1\cr \nu_2\cr}
\eqno(0.4)$$
Combining \(0.3) and \(0.4), we find that the branching matrix $B$ is
parametrized
by one parameter $t$ (which will be called scattering coefficient):
$$B=\pmatrix{1-t{2\nu_2\o \nu_1+\nu_2} &   t{2\nu_1\o \nu_1+\nu_2}\cr
               t{2\nu_2\o \nu_1+\nu_2} & 1-t{2\nu_1\o \nu_1+\nu_2}\cr}
\eqno(0.5)$$
The range of the scattering coefficient $t$ is limited by energy
conservation. The incoming energy flux is given by
$$
W_{in}={1\o 2e} (\mu_1 I_1 +\mu_2 I_2)={h\o 2e^2}(\nu_1^{-1}I_1^2
+\nu_2^{-1}I_2^2)
\eqno(0.5)$$
while the outgoing energy flux satisfies
$$W_{out}\geq {h\o 2e^2}(\nu_1^{-1}I_3^2
+\nu_2^{-1}I_4^2)
\eqno(0.6)$$
The inequality is due to the fact that the outgoing branches may not be in
equilibrium states. The equality holds only when the edges are in equilibrium
states. The condition $W_{in}=W_{out}$ (for balistic transport) or
$W_{in}>W_{out}$ (for more general cases) require $0\leq t \leq 1$.
$t=0$ represents minimum scattering (or minimum
tunneling) and $t=1$ maximum
 scattering (or maximum tunneling).
In both limits the outgoing branches are also in equilibrium states.

When the outgoing edges are in contact with voltage leads, the electrons
on the outgoing edges will reach equilibrium. The chemical potentials
measured by the voltage leads will be given by
$$\mu_3={I_3h\o \nu_1 e},\ \ \ \ \ \ \ \ \ \
\mu_4={I_4h\o \nu_2 e}
\eqno(0.7)$$
When $\nu_1\neq \nu_2$ (assume $\nu_1 >\nu_2$) and when $t$ is close to $1$
(strong scattering),
we find that $\mu_4$ can be higher than max$(\mu_1,\mu_2)$ or lower than
min$(\mu_1,\mu_2)$.
This result is quite surprising, but nevertheless, we believe, is possible
to be realized in experiments. We have plotted $\mu_{3,4}$ as a function of
$t$ in Fig. 3. We have chosen $\nu_1=1/3$ and $\nu_2=1/15$. In this case
the two edge branches can be viewed as the two edge branches of $\nu=2/5$ FQH
state.

The single impurity scattering between two edge branches of $2/5$ FQH
liquid is special since the impurity scattering is marginal at low
energies (see next section for details).
As a consequence the scattering coefficient
$t$ is independent of temperature $T$ and voltage
$\De V=(\mu_1-\mu_2)/e$ at low temperatures and low voltages. In general
the single impurity scattering can be relevant or irrelevant at low
energies. The scattering coefficient $t$ will approach $1$ (or 0) as $T$
and
$\De V$ approach to 0 if the impurity scattering is relevant (or
irrelevant). Near these fixed points, $1-t$ or $t$
will have a power law dependence on $T$ or $\De
V$.  Some
detailed discussions about the scaling of $t$ and $1-t$ near the fixed
points $t=0$ and $t=1$ can be found
in \Ref{tun,edgere} and \Ref{KF}.

\head{3. Scattering between chiral Luttinger liquids
 through a single impurity
-- microscopic considerations}

As an example, first let us consider the scattering between the two
branches of edge states of filling fraction $\nu=2/5$ FQH liquid.
One branch of the edge states corresponds to the edge states of $1/3$
Laughlin state, the other branch comes from the $\nu=1/15$ quasiparticle
condensate.\refto{edgere}
The dynamics of the two branches is determined by the following
algebra
$$\eqalign{
[\r_{1k},\r_{1k'}]=& {\nu_1\o 2\pi}  k\de_{kk'} \cr
[\r_{2k},\r_{2k'}]=& {\nu_2\o 2\pi} k\de_{kk'} \cr}
\eqno(1.1)$$
$$H=\sum_{i,j=1,2} V_{ij} \r_{i,k}\r_{j,-k}
\eqno(1.2)$$
where $(\nu_1,\nu_2)=(1/3,1/15)$ and $\r_{1,2}$ are the edge densities
of the two branches. $V_{ij}$ contains interactions between the two
branches and determine the velocities of the edge modes.
The two branches of the $2/5$ edge states propagate in the same direction
and $(V_{ij})$ is a positive definite matrix.\refto{edgere}

In presence of a single impurity, a charge $1/3$ quasiparticle may be
scattered from one branch to the other.\refto{edgere}
Such a scattering process
is described by Hamiltonian
$$H_{sc}=\Ga(e^{i(\phi_1-{\nu_1\o \nu_2}\phi_2)}+H.C.)
\eqno(1.3)$$
with ${1\o 2\pi}\part_x \phi_i(x)=\r_i(x)$.
The operator $e^{-i{\nu_1\o \nu_2}\phi_2}$ removes
$\nu_2{\nu_1\o \nu_2}$ charges from the second branch and
$e^{i\phi_1}$ adds $\nu_1$ charges to the first branch.

The correlation function of the tunneling Hamiltonian has a form
$$\<H_{sc}(t) H_{sc}(0)\>\propto {1\o t^2}
\eqno(1.4)$$
regardless the values of $V_{ij}$.\refto{edgere}
\(1.4) is identical to the correlation of the tunneling operators
between two Fermi liquids. The partition function (in Keldysh
formalism\refto{Kel})
for the scattering between two edge branches of $2/5$ FQH state
$$Z_K=\<\phi|\phi\>, \ \ \ \ \ \ |\phi\>=\cT e^{-i\int dt H_{sc}(t)}|0\>
\eqno(1.5)$$
can be written as a partition function of a 1D Coulumb gas.\refto{cg,KF,CW}
The partition function
for the scattering between two branches of 1D chiral Fermi liquids
generates the identical Coulomb gas. Thus the physical properties
of the tunneling between two edge branches of $2/5$ FQH state are
directly related
to those between two Fermi liquids.

The above result can be seen more clearly if we assume
$$(V_{ij})=\pmatrix{\nu_1^{-1}&0\cr 0&\nu_2^{-1}\cr}
\eqno(1.6)$$
In this case the two edge branches of the $2/5$ state have the same
velocity. Introducing $\t\r_{1,2}$ through
$$\eqalign{
\r_1=& \sqrt{\nu_1} (\cos(\th) \t \r_1+\sin(\th) \t \r_2)  \cr
\r_2=& \sqrt{\nu_2} (\cos(\th) \t \r_2-\sin(\th) \t \r_1)  \cr}
\eqno(1.7)$$
where
$$\tan(\th)={\sqrt{\nu_1} - \sqrt{\nu_2} \o \sqrt{\nu_1} + \sqrt{\nu_2} },
\eqno(1.7a)$$
we can rewrite \(1.1)--\(1.3) as
$$\eqalign{
[\t \r_{1k},\t \r_{1k'}]=& {1\o 2\pi} k\de_{kk'} \cr
[\t \r_{2k},\t \r_{2k'}]=& {1\o 2\pi} k\de_{kk'} \cr
H=& \sum_{i,j=1,2} 2\pi v\t \r_{i,k}\t \r_{j,-k}  \cr
H_{sc}=&\Ga(e^{i(\t \phi_1-\t \phi_2)}+H.C.)  \cr}
\eqno(1.8)$$
The coupling to the external electromagnetic field becomes
$$\eqalign{
 & eA_0^{(1)}\r_1+eA_0^{(2)}\r_2 \cr
=& e(A_0^{(1)}\sqrt{\nu_1} \cos(\th)
-A_0^{(2)} \sqrt{\nu_2} \sin(\th) )\t \r_1 +
e(A_0^{(2)}\sqrt{\nu_2} \cos(\th)
+A_0^{(1)} \sqrt{\nu_1} \sin(\th) )\t \r_2   \cr
=& e\t A_0^{(1)}\t \r_1+e\t A_0^{(2)}\t \r_2 \cr}
\eqno(1.9)$$
where $A^{(1,2)}_0$ are scalar electromagnetic potentials on the two edge
branches.
The system \(1.8) can be fermionized. Introducing effective fermion
fields
$\t\psi_i=e^{i\t \phi_i}$, we find
\(1.8) can be written as
$$
H+H_{sc}=i\sum_{i=1,2}v \t \psi_i^\dag \part_x\t \psi_i +
(\Ga \t \psi_1^\dag \t \psi_2+H.C.)
\eqno(1.10)$$
Now $H_{sc}$
describes tunneling of the effective fermions
between two (chiral) 1D free fermion systems.
$\t A^{(1,2)}_0$ in \(1.9)
are effective potentials seen by the two effective
fermions $\t \psi_{1,2}$, since $\t\r_i=\t\psi^\dag_i\psi_i$ are densities
of the effective fermions.

Due to the above mapping, many results for tunneling between two Fermi
liquids can be applied to the tunneling between two edge branches of
$2/5$ state. For example, the tunneling between two Fermi liquids is
described by the scattering coefficient $t$
$$\pmatrix{\t I_3\cr \t I_4\cr}=\pmatrix{1-t&t\cr t&1-t\cr}
\pmatrix{\t I_1\cr\t  I_2\cr}
\eqno(1.11)$$
where $\t I_{1,2}$ are the fermion currents  in the two incoming branches
and $\t I_{3,4}$ the fermion currents in the outgoing branches. Since
all excitations propagate with the same velocity, we can use \(1.7) to
relate the fermion currents $\t I_a$ to the currents on the original edge
branches
of the $2/5$ FQH liquid $I_a$:
$$
\eqalign{
I_1=& \sqrt{\nu_1} (\cos(\th) \t I_1+\sin(\th) \t I_2)  \cr
I_2=& \sqrt{\nu_2} (\cos(\th) \t I_2-\sin(\th) \t I_1)  \cr}
\eqno(1.12)$$
The above relation reduces \(1.11) to \(0.2) with $B$ precisely given by
\(0.5).
We also know, from Fermi liquid theory,
 that the parameter $t$ is independent
of voltage between the edge branches $\De V=(\mu_2-\mu_1)/e$ and
temperature $T$, if both $\De V$ and $T$ are small enough.

We can also obtain the power spectrum of the tunneling current $I_t$
from the results of free electrons in \Ref{Lev}. The noise  power
spectrum for tunneling current between two Fermi liquids is given by
$$S_{\om}= 2{e^{*2}\o \pi}\left[
t|\om|+t(1-t)(\om_J-|\om|)\th(\om_J-|\om|)  \right]
\eqno(1.13)$$
where $e^*$ is the charge transfer for each tunneling event. In our case
$e^*=\nu_1 e$. $\om_J\hbar$ in \(1.13) represent the energy increase or
decrease of the effective fermion before and after the tunneling.
Thus $\om_J$ is given by
$$\om_J=e|\t A^{(2)}_0-\t A^{(1)}_0|/\hbar={e^*\o e} |\mu_2-\mu_1|/\hbar
\eqno(1.14)$$
where we have used the result in \(1.9). Because of the mapping
discussed above, \(1.13) also applies to the tunneling between two edge
branches of $2/5$ FQH liquids. We see that  the fractional charge
$e^*=e/3$
explicitly appear in the noise spectrum. Measuring the noise spectrum
allow us to experimentally
determine the fractional charges in FQH liquids.\refto{tun,edgere}

We would like to remark that although the mapping between the $\nu=2/5$
edge states and the free Fermi liquids is valid only for special $V_{ij}$ in
\(1.6),  our results, such as \(1.13), are valid beyond the assumption
\(1.6). For general $V_{ij}$, although we can no longer map the $2/5$ edge
states to free chiral Fermi liquids as $1+1$ dimensional systems, the
Coulomb gases that describe the tunneling in the two systems are still
identical. It is because of this mapping between the Coulomb
gases\refto{CW} that the
results for tunneling between Fermi liquid can be applied to tunneling between
$2/5$ edge states. More precisely all $n$-points
correlations between tunneling currents are identical in the two systems.
 Certainly, in order for this mapping to work,
we need to do some proper translations
(such as $e\to e^*$ and the
mapping of the voltages \(1.9)).

\head{4. Scattering between two edge branches of $\nu=2$ IQH state by many
impurities}

In this section we are going to study the scattering between two
 chiral 1D Fermi liquids by many impurities. Actually we will assume a finite
density of the impurities. Such a system appears in $\nu=2$ integral QH
(IQH) state (See Fig. 4) and
$\nu=2/5$ state if we use the mapping discussed in the last section.

We know usual 1D electron gas contain two branches,
one right mover and one left mover,
corresponding to two Fermi points at $\pm k_F$. The impurities in this case
are relevant and localize the electrons. In this section we will consider
a different problem of
scattering between two branches that move in the same direction. The
impurities are still relevant at low energies but they
do not localize the motion of electrons.

In the following
we will ignore the interactions between electrons, and hence we can
consider the scattering of a single electron by the impurities.
Consider  a plane wave of an electron of energy $\eps$. Let the amplitudes of
the
plane wave to be
$(A_1,A_2)$ in the two branches at the point $A$ in Fig. 4. After passing
through the disordered region the electron has an amplitude $(B_1,B_2)$ at
point $B$ in Fig. 4. The amplitudes at $A$ and $B$ are related by
a unitary transfer matrix $T$
$$\pmatrix{B_1\cr B_2\cr}=
U_NP_{N-1}U_{N-1}...P_1U_1\pmatrix{A_1\cr A_2\cr}
\equiv T\pmatrix{A_1\cr A_2\cr}
\eqno(2.1)$$
In \(2.1) $U_i$ is a two by two matrix describing the scattering of the
i$^{th}$ impurity. Here we will consider a simple model that $U_i$
takes the following form
$$U_i=e^{i\th_i}W_i
\eqno(2.2)$$
where $\th_i$ is a random phase, and
$W_i$ is a random $SU(2)$ matrix with a uniform (\ie $SU(2)$ invariant)
distribution. The matrix $P_{i}$ describes
 the propagation between the i$^{th}$ and (i+1)$^{th}$ impurities and has
the form
$$P_i=\pmatrix{e^{i\eps l_i/v_1}&0\cr 0&e^{i\eps l_i/v_2}\cr}
=e^{i\eps l_i/\bar v}e^{i\eps l_i {1\o 2}(v_1^{-1}-v_2^{-1})\si_3 }
\eqno(2.3)$$
where $l_i$ is the
separation between the i$^{th}$ and (i+1)$^{th}$ impurities,
$v_{1,2}$ are the velocities in the two branches and
$\bar v=2v_1v_2/(v_1+v_2)$ is an average of the two velocities.
The $\si_3$ term in \(2.3) belongs to $SU(2)$ group
and can be absorbed in $W_i$.
Thus we can drop the $\si_3$ term and assume
$$P_i=e^{i\eps l_i/\bar v}
\eqno(2.4)$$
In this case the transfer matrix $T$ has a form
$$T=e^{i\eps l_{AB}/\bar v}e^{i\sum\th_i}W_N...W_1
\eqno(2.5)$$
We see the difference of the two velocity $v_1^{-1}-v_2^{-1}$ becames
unimportant in our model due to the impurity effect. As an consequence we
expect that {\it
any} disturbance will propagate roughly with the same velocity $\bar v$ in
the disordered region even though the two branches may originally have very
different velocities. (More detailed discussion will be given later.)

Certainly $W_i$ and $\th_i$ are also functions of $\eps$. At a new energy
$\eps+\De \eps$, $W_i$ and $\th_i$ will change into
$$W'_i=e^{i\vec \phi'_i\cdot \vec \si}W_i,\ \ \ \ \ \ \
\th'_i=\th_i+\De \th_i
\eqno(2.6)$$
$\vec\phi_i'$ and $\De\th_i$ in \(2.6) are random variables. $|\vec\phi_i'|$
 and $|\De\th_i|$ are proportional to $\De\eps$ if $\De\eps$ is small
enough. Note here we do not assume $\vec\phi_i'$ to have a uniform
distribution in all directions. For example $\vec\phi_i'$  may arise from
the propagator $P_i$ in \(2.3). In this case
$$\vec\phi_i'\cdot\vec\si=\De\eps{l_i\o 2}(v_1^{-1}-v_2^{-1})\si_3
$$
The transfer matrix at energy $\eps+\De \eps$ becomes
$$T(\eps+\De \eps)=e^{i\sum\De\th_i}e^{i\vec \phi_N\cdot \vec \si}
...e^{i\vec \phi_1\cdot \vec \si}T(\eps)
\eqno(2.7)$$
where
$$e^{i\vec \phi_i\cdot \vec \si}=Ue^{i\vec \phi'_i\cdot \vec \si}U^{-1},
\ \ \ \ \ \ \
U=W_N...W_{i+1}
\eqno(2.8)$$
Since $U$ is random, $\vec \phi_i$ will have a random direction with
uniform distribution (but with
same length as $\vec \phi'_i$). The directions of $\vec \phi_i$ and
$\vec \phi_{i-1}$ are uncorrelated since $W_i$ is a random $SU(2)$ matrix.

{}From the above discussions we can obtain the following results.
At a given energy, $T$ is a random matrix in $U(2)$ group with a uniform
probability distribution over the group manifold. The ratio of the
transfer matrix at two different energies are generated by brownian
motions, one on the group manifold $U(1)$ and the other on $SU(2)$. More
precisely, we have
$$T(\eps+\De \eps)T^{-1}(\eps)=e^{i\sum\De\th_i}
   e^{i\vec \phi_N\cdot \vec \si}
...e^{i\vec \phi_1\cdot \vec \si}
\eqno(2.9)$$
where $(\De\th_1,\De\th_2,...)$ has a probability distribution
$$P(\De\th_1,\De\th_2,...)=\prod_i P_\th(\De\th_i)
\eqno(2.10)$$
and $(\vec \phi_1,\vec \phi_2,...)$ has a probability distribution
$$P(\vec \phi_1,\vec \phi_2,...) =\prod_i P_\phi(|\vec \phi_i|)
\eqno(2.11)$$
where $P_\th$ and $P_\phi$ are probability distributions of
single-variables $\De\th_i$ and $\vec\phi_i$.

To discuss the probability distribution of
$$U(\De \eps)\equiv T(\eps+\De \eps)T^{-1}(\eps)\equiv e^{i\De\Th}
e^{i\vec \Phi\cdot \vec \si}
\eqno(2.12)$$
It is convenient to introduce
$$\eqalign{
\De \bar\th=& \sqrt{ \int d\De \th (\De\th)^2 P_\th(\De\th) }, \cr
\bar\phi=& \sqrt{ {1\o 3}\int d^3\vec\phi (\vec\phi)^2 P_\phi(|\vec \phi|)
}.\cr}
\eqno(2.13)$$
Note for small $\De\eps$, $\De \bar\th$ and $\bar\phi$ are proportional to
$\De\eps$.
The probability distribution of the phase of $U(\De \eps)$ in \(2.12),
$e^{i\De\Th}$,  is described by
the brownian motion on a circle. $\De \bar\th$ is the average
length of each step. Thus
for large $N$, $e^{i\De\Th}$ has a distribution
$$P(e^{i \De\Th})=(2\pi N\De\bar \th^2)^{-1/2} \sum_{n=-\infty}^\infty
 e^{-{(\De\Th+2n\pi)^2\o 2N (\De \bar\th)^2}}
\eqno(2.14)$$
We see that the distribution of the $U(1)$ phase of $U$ in \(2.12)
is an universal function which depends only on $\sqrt{N}\De \bar\th$.

The probability distribution of $e^{i\vec \Phi\cdot \vec \si}$ is described
by the brownian motion on the $SU(2)$ group manifold and is more
complicated. However it is also an universal function which
depends only on $\sqrt{N}\bar\phi$ as long as $\bar\phi \ll 1$.
When $\sqrt{N}\bar\phi\ll 1$, $e^{i\vec \Phi\cdot \vec \si}$ has an
universal Gaussian distribution because $\vec\phi_i$ can be added linearly
in \(2.9):
$$P(e^{i\vec \Phi\cdot \vec \si})=(2\pi N\bar \phi^2)^{-3/2}
e^{-{ (\vec\Phi)^2\o 2N \bar\phi^2} }
\eqno(2.15)$$

To calculate the distribution for more general cases, we notice that
the $SU(2)$ group manifold is parameterized by a 3D vector $\vec\Phi$
with $\Phi=|\vec\Phi|<\pi$, through the relation
$W=e^{i\vec\Phi\cdot\vec\si}$. The $SU(2)$ invarient volume element
 is given by
$$dU=\sin^2(\Phi) d\Phi d\Om
\eqno(2.15a)$$
where $d\Om$ is the solid angle of the vector $\vec\Phi$. The  probability
distribution $P(e^{i\vec \Phi\cdot \vec \si})$ is a solution of the
diffusion equation
$$ (D\part_t+H)P(\vec\Phi,t)=0
\eqno(2.15b)$$
at $t=1$, \ie $P(e^{i\vec \Phi\cdot \vec \si})=P(\vec\Phi,t=1)$. The
initial condition at $t=0$ is taken to be
$P(\vec\Phi,t=0)=\de^3(\vec\Phi)$. $-H$ in \(2.15b) is the Laplacian on
the $SU(2)$ group manifold
$$H=-{1\o \sin^2(\Phi)}\part_\Phi  \sin^2(\Phi) \part_\Phi+
{\vec L^2\o  \sin^2(\Phi) }
\eqno(2.15c)$$
where $\vec L$ is the angular momentum operator for the rotation of
$\vec\Phi$. The coefficient $D$ is given by
$$D={2\o N \bar\phi^2}
$$
In the $\vec L=0$ secter the normalized eigenfunction of $H$ is given by
$$\chi_n(\Phi)={1\o\sqrt{2}\pi} {\sin(n\Phi)\o \sin(\Phi)}, \ \ \ \ \
n=1,2,3...
\eqno(2.15d)$$
with eigenvalue $n^2-1$. $P(\vec\Phi,t)$ is then given by
$$P(\vec\Phi,t)=\sum_{n=1}^\infty \chi_n(0)\chi_n(\Phi) e^{-(n^2-1)t/D}
\eqno(2.15e)$$
Thus we find
$$P(e^{i\vec \Phi\cdot \vec \si})
={1\o 2\pi^2}\sum_{n=1}^\infty n{\sin(n\Phi)\o \sin(\Phi)} e^{-{1\o
2}(n^2-1)N\bar\phi^2}
\eqno(2.15f)$$
We see that the distribution indeed depends only on $\sqrt{N}\bar\phi$.
{}From \(2.14) and \(2.15f), we can also obtain the joint probability
distribution for two transfer matrices at two different energies
$$P(T(\eps),T(\eps+\De\eps))=
{1\o 4\pi^3}P(e^{i \De\Th})P(e^{i\vec \Phi\cdot \vec \si})
\eqno(2.15g)$$
where $\De\Th$ and $\vec\Phi$ are given in \(2.12).

We see that the transfer matrix $T$ will has a significant change of
order 1 if
we change $\eps$ by $\De\eps_c \sim \ga_\phi/\sqrt{N} $, where
$\ga_\phi=\De\eps/\bar\phi$ which approaches to constant for small
$\De\eps$. From the energy scale $\De\eps_c$ and the uniform distribution
of the transfer matrix $T$, we can immediately obtain the following
results. When the temperature and voltage difference between the two
branches are much less than $\De\eps_c $, the
branching matrix $B$ for the current will fluctuate strongly as we change the
chemical potential on the edges. However if the temperature or the voltage
 difference between the two
branches is much larger than $\De\eps_c $, the branching matrix $B$ in
\(0.2) will
have the following universal form
$$B={1\o 2} \pmatrix{1&1\cr1&1\cr}
\eqno(2.16)$$

When $|\mu_1-\mu_2| \ll \De\eps_c$
we are in the linear response regime and the impurity scattering is
described by the branching matrix $B$ in
\(0.5) with $\nu_1=\nu_2=1$. The scattering coefficient $t$ can
be calculated from the transfer matrix $T$. At finite temperatures we
have
$$t(\mu)=\int d\eps f_F(\eps-\mu)|T_{12}(\eps)|^2
\eqno(2.16a)$$
where $f_F(\mu)=-dn_F(\mu)/d\mu$ is the derivative of the
Fermi distribution function and $\mu$ is the average
chemical potential on the two incoming branches. At the zero temperature
$t(\mu)$ fluctuates randomly between 0 and 1 as we change the chemical
potential $\mu$. The probability distribution of the scattering
coefficient $t=1-|T_{11}|^2$ is given by
$$
P(t)= {1\o 2\pi^2}\int \sin^2(\Phi)d\Phi \sin(\th) d\th d\phi
      \de[t-1+ \cos^2(\Phi)+\cos^2(\th)\sin^2(\Phi)]
    =1
\eqno(2.16aa)$$
However $t(\mu)$ can have a significant change only when we change
$\mu$ by an amount of order $\De \eps_c$. We would like to point out
that the constant probability distribution $P(t)=1$ is a consequency of the
uniform distribution of the transfer matrix $T$ over the $SU(2)$ group
manifold. We expect to have a different distribution $P(t)$ if the transfer
matrix has a non-uniform distribution or if there are more than two branches of
edge excitations that are involved in the impurity scattering.

At temperatures $T>\De \eps_c$,
the fluctuations of $t(\mu)$ around $t=1/2$ will be suppressed. For $T\gg \De
\eps_c$, $t(\mu)$ approaches to a universal value $1/2$. Using the
joint probability distribution \(2.15g) we can also calculate the exact
correlation $\< t(\mu_1)t(\mu_2)\>$ for two scattering coefficients
at two different chemical potentials:
$$\< t(\mu_1)t(\mu_2)\>=
\int d\eps d\eps'
f_F(\eps-\mu_1)|T_{12}(\eps)|^2
f_F(\eps'-\mu_2)|T_{12}(\eps')|^2
P(T(\eps),T(\eps'))
\eqno(2.16b)$$
which depends only on $\mu_1-\mu_2$.

The time it takes for a wave packet to propagate from $A$ to $B$ is given
by, for example, $d\ln(T_{11}(\eps))/d\eps$ which is close to $L/\bar
v$. The fluctuations of the propagation time due to the randomness are
proportional to $\sqrt{N}$ since
the distribution of
$T(\eps)T^\dag(\eps+\De\eps)$ is a function of $\De\eps\sqrt{N}$.
Thus the uncertainty of the velocity $\De v_\eps/\bar v$ is proportional
to
$1/\sqrt{N}$ which is small in large $N$ limit.
This result is easy to understand in our model \(2.1).
The electron on average spend half time on the first branch and the other
half on the second
branch regardless the initial condition. Thus all excitation propagate
with the same average velocity $\bar v$. The $1/\sqrt{N}$ fluctuation
is due to the randomness.

We also like to stress that the probability distribution of the transfer
matrix $T(\eps)$ satisfy an $SU(2)\times SU(2)=O(4)$ symmetry. For example
the joint probability distribution $P(T_1,T_2)$ in \(2.15g) is invariant
under the transformation $T_{1,2}\to W T_{1,2} U$ where $W,U \in SU(2)$.
This is remarkable since the clean model, having different velocities in the
two branches, has no such $SU(2)\times SU(2)$ symmetry. The impurity
scattering is relevant at low energies. Our results indicate that
the impurities drive the system to a new fixed
point which has  the additional $SU(2)\times SU(2)$
symmetry. We note that the clean system can have an $SU(2)$
symmetry if the velocity in the two branches are the same. However the new
disordered fix point is different from the $SU(2)$ symmetric
clean system, as we can see from the correlation functions calculated in
the next section.

The above results can be generalized to problem of impurity scattering
between $n$ branches of chiral 1D Fermi liquids. The $n\times n$ transfer
matrix will have a uniform distribution over the $U(n)$ group manifold.
The ratio of two transfer matrices at two different energies
$T(\eps)T^{-1}(\eps+\De\eps)$ is again generated by brownian motions on
the $U(1)$ and $SU(n)$
group manifolds, and the probability distribution can be calculated from
the diffusion equations on the $SU(n)$ group manifold.

In the above we have calculated the joint probability distribution of two
transfer matrices at different energies. In experiments the transfer matrix
may also depend on other parameters, such as magnetic field $B$. We can use
the same approach to calculate the joint probability distribution of two
transfer matrices at different magnetic fields and obtain the same result
as in \(2.15g):
$$P(T(\eps,B),T(\eps+\De\eps, B+\De B))=
{1\o 4\pi^3}P(e^{i \De\Th})P(e^{i\vec \Phi\cdot \vec \si})
\eqno(2.16c)$$
 But now $\De\bar\th$ and $\bar\phi$ in \(2.14) and \(2.15f)
are parameters that depend on $\De\eps$ and $\De B$. For small $\De\eps$
and $\De B$ they have the form
$$\eqalign{
\De\bar\th^2=& a \De\eps^2 +b \De B^2 + c \De\eps\De B \cr
\bar\phi^2=& a' \De\eps^2 +b' \De B^2 + c' \De\eps\De B \cr}
$$
The probability distribution of the scattering
coefficient $t$ is still given by $P(t)=1$.  The correlation
$\< t(\mu, B_1)t(\mu, B_2)\>$ for two scattering coefficients
at two different parameters $B$ will be given by
$$\eqalign{
 &\< t(\mu,B_1)t(\mu,B_2)\> \cr
=&\int d\eps d\eps'
f_F(\eps-\mu)|T_{12}(\eps,B_1)|^2
f_F(\eps'-\mu)|T_{12}(\eps',B_2)|^2
P(T(\eps,B_1),T(\eps',B_2))      \cr}
\eqno(2.16d)$$

\head{5. Correlation functions}

In this section we are going to use the transfer matrix to calculate some
correlation functions along the disordered edge. Let $T(\eps,x)$ be
the transfer matrix from $x=0$ to $x$ at energy $\eps$. The transfer
matrix from $x$ to $y$ will be given by $T(\eps,y)T^\dag(\eps,x)$.
The transfer matrix calculated in last section $T(\eps,N)$
depend on $\eps$ and $N$.
The two transfer matrices are related through
$$T(\eps,x)=T(\eps,N=|x|/d)
\eqno(3.1)$$
where $d$ is the mean separation between impurities in our model \(2.1).
The physical meaning of the length scale $d$ is that the transfer matrix
will become significantly different
from 1 when $|x|\sim d$. (Note $T(\eps,x=0)=1$.)
Only when $x> d$ does the transfer matrix have uniform probability
distribution. With the identification
$$N=|x|/d
\eqno(3.2)$$
we can also obtain the joint probability distribution:
$$P[T(\eps,x),T(\eps+\De\eps,x)]=
{1\o 4\pi^3}P(e^{i \De\Th},|x|\De\eps^2\eta_1^2)P(e^{i\vec \Phi\cdot \vec
\si},|x|\De\eps^2\eta_2^2)
\eqno(3.3)$$
where
$$P(e^{i \De\Th},\a)=(2\pi N\De\bar \th^2)^{-1/2} \sum_{n=-\infty}^\infty
 e^{-{(\De\Th+2n\pi)^2/2\a}}
\eqno(3.4)$$
and
$$P(e^{i\vec \Phi\cdot \vec \si},\a)
={1\o 2\pi^2}\sum_{n=1} n{\sin(n\Phi)\o \sin(\Phi)} e^{-{1\o 2}(n^2-1)\a}
\eqno(3.5)$$
$\De\Th$ and $\vec\Phi$ is again given by \(2.12). $\eta_{1,2}$ in \(3.3)
are given by
$$\eqalign{
\eta_1=&{\De \bar \th \o\De\eps \sqrt{d}} \cr
\eta_2=&{\bar \phi\o\De\eps \sqrt{d}} \cr}
\eqno(3.5a)$$
which approach to constants for small $\De\eps$.

Let $(c_{1\eps}^\dag,c_{2\eps}^\dag)$ be the electron operators that
create the two orthogonal states of energy $\eps$. In terms of the
transfer matrix, we have
$$\pmatrix{c_{1\eps}\cr c_{2\eps}\cr}=
\int dx T^\dag(\eps,x) \pmatrix{c_{1}(x)\cr c_{2}(x)\cr}
\eqno(3.6)$$
The inverse is given by
$$\pmatrix{c_{1}(x)\cr c_{2}(x)\cr}=
\int {d\eps\o 2\pi v_\eps} T(\eps,x) \pmatrix{c_{1\eps}\cr c_{2\eps}\cr}
\eqno(3.7)$$
where $v_\eps$ is the velocity at energy $\eps$ which can be taken to be
the
constant $\bar v$ since the fluctuation of $v_\eps$ is small.
($\De v_\eps/v_\eps \propto 1/\sqrt{L}$ where $L$ is the length of the
edge.)

{}From \(3.7) we obtain the electron propagator along the edge:
$$\eqalign{
\<c_\a^\dag(x_1,t_1) c_\be(x_2,t_2)\>
=& \int_{-\infty}^0 {d\eps\o 2\pi v_\eps}
[T(\eps,x_1) T^\dag(\eps,x_2)]^*_{\a\be}
e^{i\eps(t_1-t_2)}  \cr
\<c_\a(x_1,t_1) c^\dag_\be(x_2,t_2)\>
=& \int_0^{+\infty} {d\eps\o 2\pi v_\eps}
[T(\eps,x_1) T^\dag(\eps,x_2)]_{\a\be}
e^{-i\eps(t_1-t_2)}  \cr}
\eqno(3.6a)$$
We can also calculate correlation function of the following
two-fermion operators $\r^\mu=c^\dag_\a \si^\mu_{\a\be}c_\be$, $\mu=0,1,2,3$
and $(\si^0,\si^i)=(1,\vec \si)$:
$$\<\r^\mu(x,t)\r^\nu(0)\>=
\int_0^{+\infty} {d\eps_1\o 2\pi v_{\eps_1}}
\int_{-\infty}^0 {d\eps_2\o 2\pi v_{\eps_2}}
\hbox{Tr} \left(\si^\mu T(\eps,x)\si^\nu T^\dag(\eps,x)\right)
e^{-i(\eps_1-\eps_2)t}
\eqno(3.7a)$$
Note $\r^0=\r_1+\r_2$ is the total density and $\r^3=\r_1-\r_2$ is the
density difference of the two edge branches. The impurity average
(denoted by $\<\< ... \>\>$) of the above correlation functions in \(3.6a)
and \(3.7a) are zero for $x\gg d$, except $\<\r^0 \r^0\>$. This is a
consequence of the $SU(2)\times SU(2)$ symmetry. The impurity averaged
correlation between the total densities can be calculated exactly:
$$\eqalign{
 & \<\<\r^0(x,t)\r^0(0)\>\>  \cr
=&
\int_0^{+\infty} {d\eps_1\o 2\pi v_{\eps_1}}
\int_{-\infty}^0 {d\eps_2\o 2\pi v_{\eps_2}}
\int_0^\pi 4\pi \sin^2(\Phi)d\Phi
\int_0^{2\pi} d\De\Th    \cr
 & 2\cos(\Phi)P(\De\Th, |x|(\eps_1-\eps_2)^2\eta_1^2)
P(\Phi,|x|(\eps_1-\eps_2)^2\eta_2^2)
e^{-i(\eps_1-\eps_2)(t-\bar v^{-1}x)}  e^{i\De\Th} \cr}
\eqno(3.8)$$
where $P(\De\Th,\a)$ and $P(\Phi,\a)$ are given by \(3.4) and \(3.5).
We see that $\<\<\r^0(x,t)\r^0(0)\>\> $ has the following form
$$\<\<\r^0(x,t)\r^0(0)\>\> ={1\o \bar v^2 |x|\eta_2^2}
f({t-\bar v^{-1}x\o \sqrt{|x|}\eta_2},{ \eta_1\o \eta_2})
\eqno(3.9)$$
where the universal function $f$ is given by
$$\eqalign{
 & f(\ga_1,\ga_2) \cr
=&
\int_0^{+\infty} {d\a_1\o 2\pi }
\int_{-\infty}^0 {d\a_2\o 2\pi }
\int_0^\pi 4\pi \sin^2(\Phi)d\Phi
\int_0^{2\pi} d\De\Th    \cr
 & 2\cos(\Phi)e^{i\De\Th} P(\De\Th, (\a_1-\a_2)^2\ga_2^2)
P(\Phi,(\a_1-\a_2)^2)
e^{-i(\a_1-\a_2)\ga_1}  \cr
=&{1\o 2\pi^2(\ga_2^2+{3\o 2})}
\int_0^\infty \a e^{-\a^2}e^{-i\ga_1\a/\sqrt{\ga_2^2+{3\o2}}}d\a  \cr
}
\eqno(3.10)$$
Thus \(3.9) can be rewritten as
$$\<\<\r^0(x,t)\r^0(0)\>\>
={1\o 2\pi^2 \bar v^2 |x|(\eta_1^2+{3\o 2}\eta_2^2)}
g({t-\bar v^{-1}x\o \sqrt{|x|}\sqrt{\eta_1^2+{3\o 2}\eta_2^2}  })
\eqno(3.9a)$$
The universal function $g(\ga)$ is given by
$$g(\ga)=\int_0^\infty \a e^{-\a^2}e^{-i\ga\a}d\a
\eqno(3.10a)$$
The universal function $g(\ga)$ has a peak at $\ga=0$. We see
that the correlation $\<\<\r^0(x,t)\r^0(0)\>\>$ has a peak at $x=\bar v
t$. The width of the peak is proportional to $\sqrt{|x|}$. Other
correlation functions, such as $\< c_\a(x,t)c_\be(0)\>$ before impurity
average,
are expected to have
the similar behavior. At finite temperatures, \(3.9a) takes a form
$$\<\<\r^0(x,t)\r^0(0)\>\>
={1\o 2\pi^2 \bar v^2 |x|(\eta_1^2+{3\o 2}\eta_2^2)}
g({t-\bar v^{-1}x\o \sqrt{|x|}\sqrt{\eta_1^2+{3\o 2}\eta_2^2}  },
T\sqrt{|x|(\eta_1^2+{3\o 2}\eta_2^2) } )
\eqno(3.9b)$$
where the finite-temperature universal function $g(\ga,\be)$ is given by
$$g(\ga,\be)=\int_{-\infty}^\infty  (1+e^{-\a_1/\be})^{-1}(1+e^{\a_2/\be})^{-1}
e^{-(\a_1-\a_2)^2}e^{-i\ga(\a_1-\a_2)}d\a_1d\a_2
\eqno(3.10b)$$

Note that the correlations between the total densities can be measured
in edge magnetoplasma experiments.\refto{Ash}
 Our results indicate that, in presence of
impurities, edge magnetoplasma experiments can see only one branch of edge
excitations despite the clean system may contain several edge branches.

\head{6. Summary}

In this paper we studied the scattering between two branches of
1D chiral Luttinger
liquids by a single impurity. We found, in the linear response regime,
the current branching matrix in \(0.2) takes the following form:
$$B=\pmatrix{1-t{2\nu_2\o \nu_1+\nu_2} &   t{2\nu_1\o \nu_1+\nu_2}\cr
               t{2\nu_2\o \nu_1+\nu_2} & 1-t{2\nu_1\o \nu_1+\nu_2}\cr}
$$
where $\nu_1$ and $\nu_2$ are conductances of the two branches (measured
in units $e^2/h$). The single-impurity scattering between two edge
branches of the $2/5$ FQH liquid was studied in more detail. We found that the
problem of tunneling between two 2/5 edge branches can be mapped to the
problem of tunneling between two chiral 1D Fermi liquids. This mapping not
only allow us to rederive the above result about the branching matrix $B$,
it also allow us to calculate the noise spectrum of the tunneling current.

We also considered the scattering between two
branches of 1D chiral Fermi liquids by many impurities (see Fig. 4).
Apart from the obvious result that the transfer matrix $T(\eps)$ is a
random matrix with a uniform distribution over the $U(2)$ group manifold,
we calculate the exact joint probability distribution of two transfer
matrices at different energies (or other parameters, such as magnetic
fields, that control the transfer matrix). This allows us to study the
behavior of the
scattering coefficient $t$ in the branching matrix as a function of the
chemical potential $\mu$, magnetic field, and other parameters. We found
that as we change, for example, the chemical  potential
$\mu$, $t$ fluctuates
randomly between 0 and 1 at zero temperature. The probability
distribution of $t$ was found to be $P(t)=1$. The
sensitivity of $t(\mu)$ to the change of $\mu$ is controlled by an
energy scale $\De\eps_c \sim {\De v \o \sqrt{x d}}$, where $\De v$ is
the difference of the Fermi velocities in the two branches,
$x$ is the length of the scattering region, and $d$ is
the length scale beyond which impurities start to induce significant
scattering between the two branches. We found that $t(\mu)$ may has an order
1 change only if $\mu$ is changed by an amount of order $\De\eps_c$. Or more
precisely, the typical magnitude of the gradient $dt(\mu)/d\mu$ is of order
$1/\De\eps_c$. For temperatures $T \gg\De\eps_c$, the scattering
coefficient $t(\mu)$ approaches to a universal value $1/2$.
An exact expression of the correlation $\< t(\mu_1) t(\mu_2)\>$ at
finite temperatures was
also obtained.

\references

\refis{edgere} For a review see
X.G. Wen, \journal Int. J. Mod. Phys., B6, 1711, 1992.

\refis{top}
X.G. Wen,
\journal Int. J. Mod. Phys. B, 2, 239, 1990;
\pr B40, 7387, 1989;
X.G. Wen and Q. Niu,  \pr B41, 9377, 1990.

\refis{BW}
B. Blok and X.G. Wen, \pr B42, 8133, 1990; \pr B42, 8145, 1990.



\refis{Halp}
B.I. Halperin, \pr B25, 2185, 1982;
Q. Niu and D.J. Thouless, \pr B35, 2188, 1987.

\refis{Ash} R.C. Ashoori \etal, \pr B45, 3894, 1992.


\refis{tun}
X.G. Wen, \pr B44, 5708, 1991.


\refis{edge}
X.G. Wen, \prb 43, 11025, 1991.

\refis{elec}
X.G. Wen, \prl 66, 802, 1991.

\refis{JM}
M.D. Johnson and A.H. MacDonald, \prl 67, 2060, 1991.

\refis{FH} H.A. Fertig and B.I. Halperin,
\pr B36, 7969, 1987.

\refis{CW} C. Chamon, D.E. Freed, and X.G. Wen, to appear.

\refis{cg} A.O. Caldeira and A.J. Leggett, \journal Ann. Phys. (N.Y.),
149, 374, 1983; A. Schmid, \prl 51, 1506, 1983;
F. Guinea, V. Hakim, and A. Muramatsu, \prl 54, 263, 1985;
F. Guinea, \pr B32, 7518, 1985.

\refis{Lev} L.S. Levitov and G.B. Lesovik, \journal JETP Lett., 58, 230, 1993;
S.-R. E. Yang, \journal Solid State Comm., 81, 375, 1992.


\refis{M}
A.H. MacDonald, \prl 64, 220, 1990.

\refis{FK}
J. Fr\"ohlich and T. Kerler, \np B254, 369, 1991.

\refis{KF}
C.~L.~Kane and M.~P.~A.~Fisher
  Phys. Rev. Lett. {\bf 68}, 1220 (1992);
 Phys. Rev. B {\bf 46}, 15233 (1992).

\refis{Kel}
L.V. Keldysh, \journal Soviet Phys. JETP, 20, 1018, 1965.

\endreferences

\vfil \eject
\head{ Figure captions}

\item{Fig. 1}
Scattering of (or tunneling between)
two chiral Luttinger liquids by an impurity. The arrows indicate
the direction of propagation. 1 and 2 (3 and 4) label the incoming (outgoing)
branches of edge 1 and edge 2 respectively.

\item{Fig. 2}
Experimental realizations of the scattering between chiral Luttinger liquids.
(a) Scattering between two branches of edge states of $\nu=2/5$ FQH liquids.
The conductances of the two branches is given by
$(\nu_1,\nu_2)e^2/h =(1/3,1/15)e^2/h$. The impurity scattering can transfer
charge $1/3$
quasiparticles in the $1/3$ FQH liquid from one branch to the other.
The impurity scattering is marginal at low energies.
(b) Scattering between two branches of edge states of $\nu=2/3$ FQH liquids.
The conductances of the two branches is given by
$(\nu_1,\nu_2)e^2/h =(1,1/3)e^2/h$. The impurity scattering can only transfer
charge $1$ electrons from one branch to the other.
The impurity scattering is irrelevant at low energies.
In the above we have ignored the interaction between the two edge branches.
Thus it is important to separate the two edge branches in
the region away from the
scattering region (as illustrated in the figure)
in order for the above results to be valid.

\item{Fig. 3}
The chemical potentials $\mu_{3,4}$ on the outgoing
edge branches for the impurity
scattering between two edge states of $2/5$ FQH liquid (Fig. 2a).
$t$ is the scattering coefficient that characterizes the strength of the
scattering.  We have
chosen $(\mu_1,\mu_2)=(0,1)$. Note for $t> 0.6$,
$\mu_4 < \hbox{min}(\mu_1,\mu_2)=0$. Also notice that $\mu_3=\mu_4$ when
$t=1/2$.

\item{Fig. 4}
Multi-impurity scattering between two 1D chiral Fermi liquids.

\vfil \eject

\epsfxsize=6truein
\epsffile{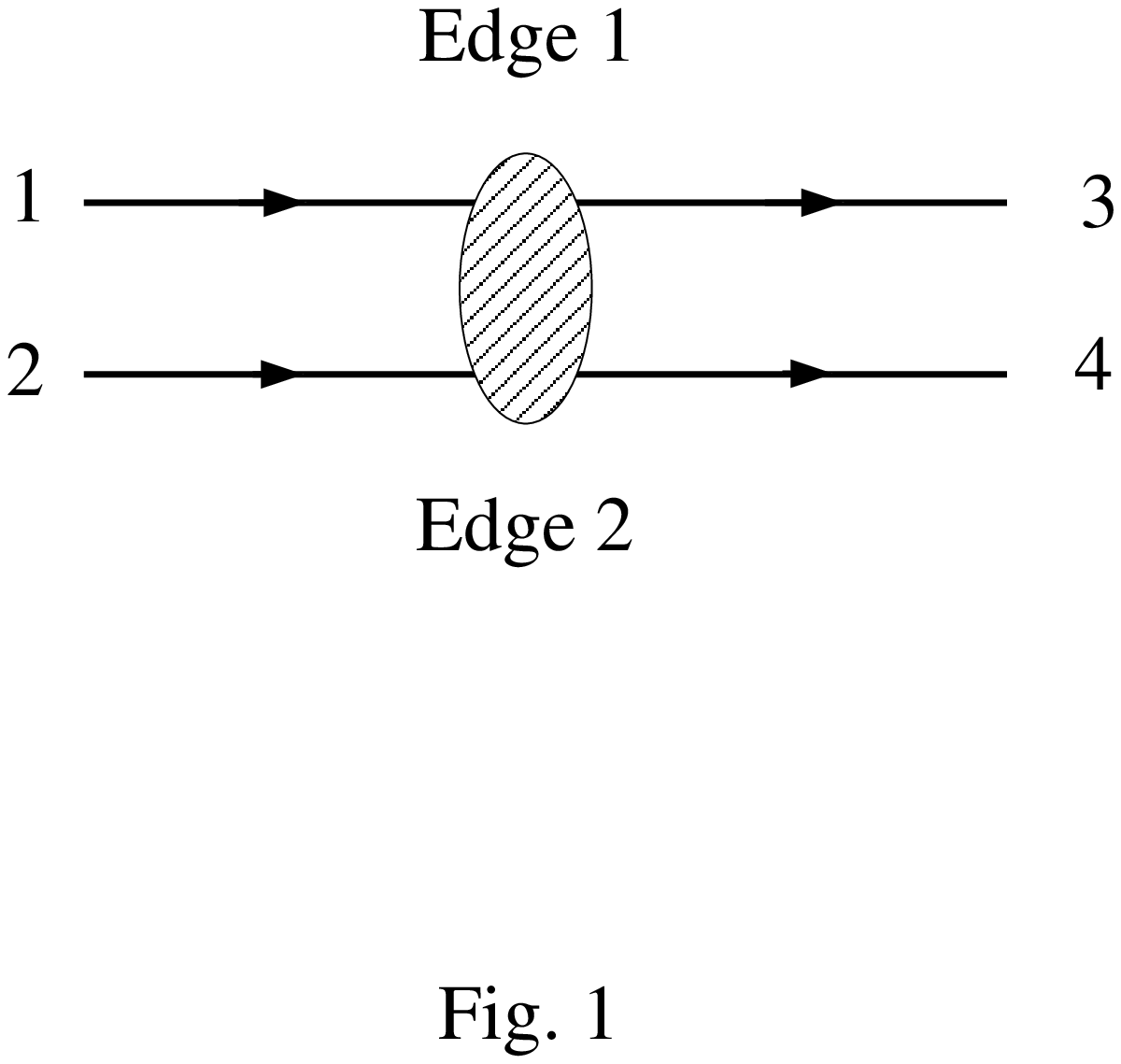}
\vfil \eject

\epsfxsize=6truein
\epsffile{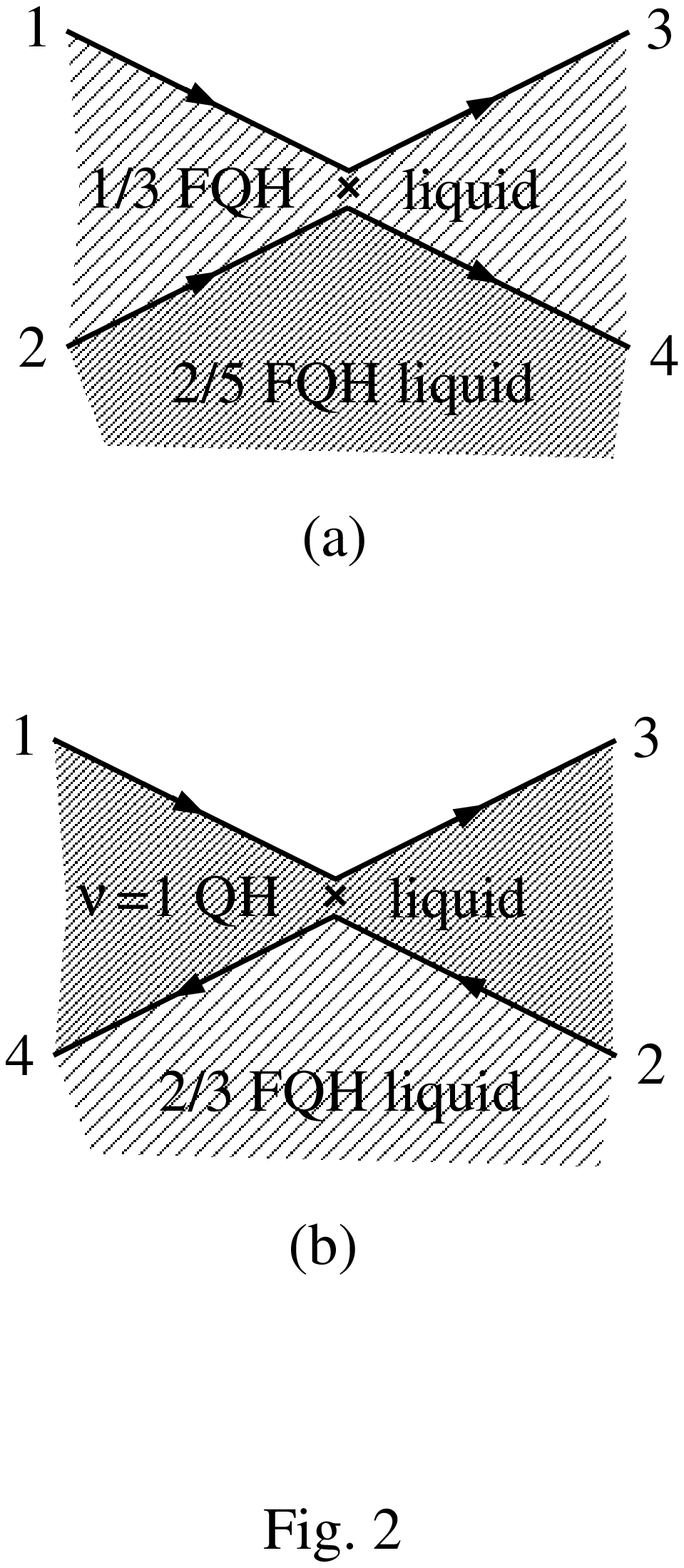}
\vfil \eject

\epsfxsize=6truein
\epsffile{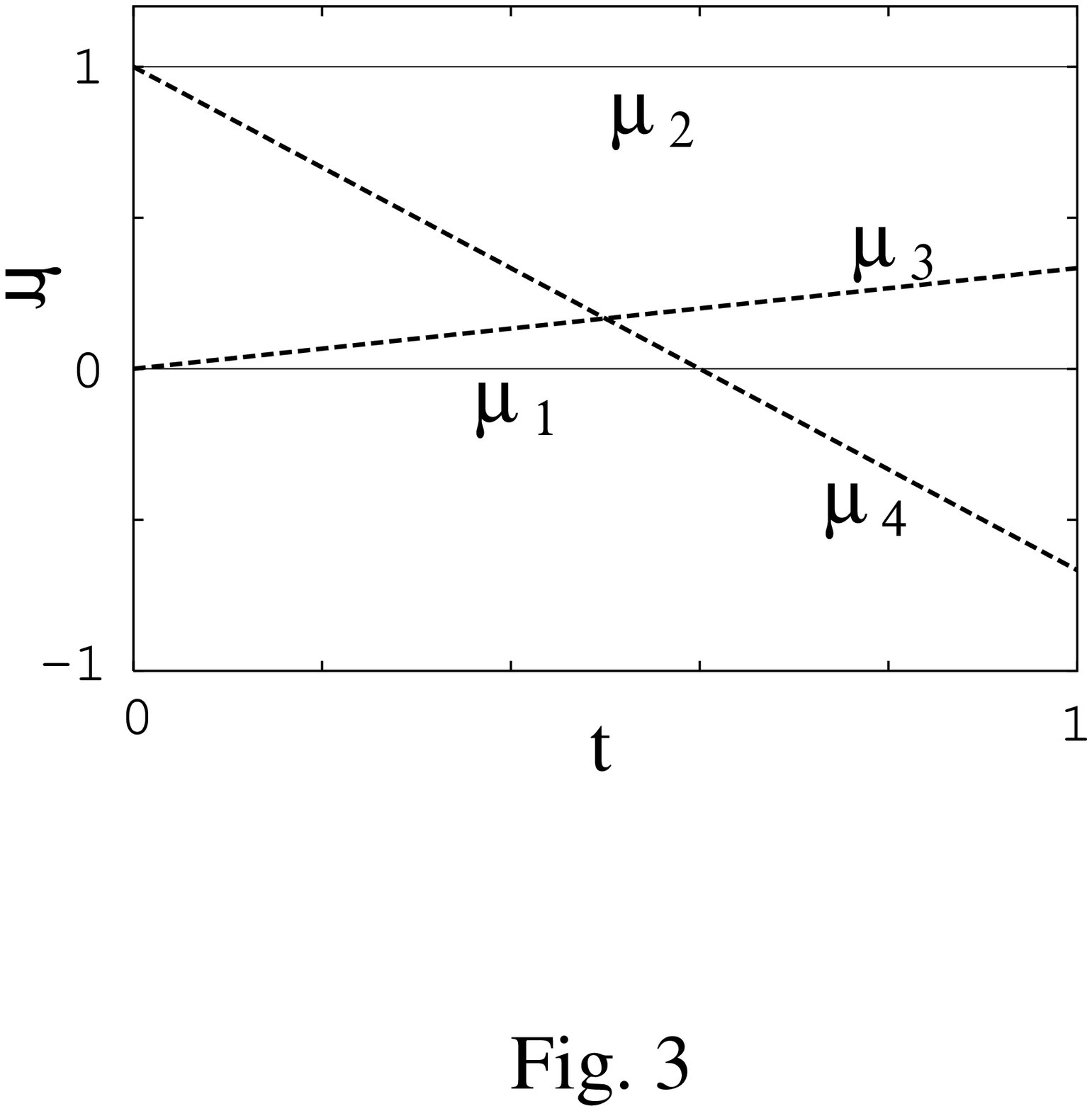}
\vfil \eject

\epsfxsize=6truein
\epsffile{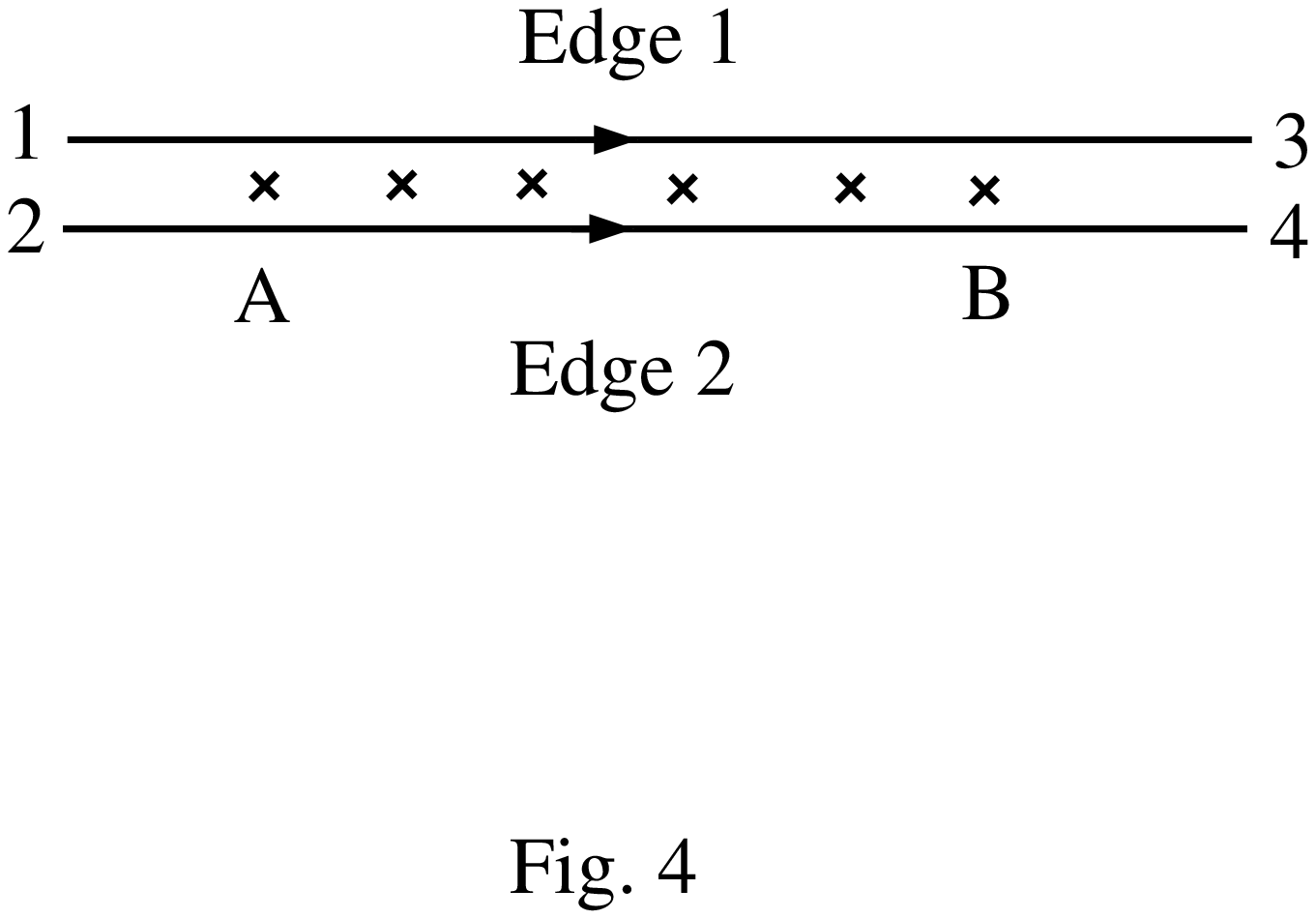}

\end